\documentclass[onecolumn,11pt, draftcls]{IEEEtran} 

\usepackage{amsmath,dsfont}
\usepackage{amssymb,amsthm}
\usepackage{epsfig,verbatim}
\usepackage{mathrsfs}
\usepackage{verbatim}
\usepackage{syntonly}
\usepackage{graphicx}
\usepackage{epstopdf}
\usepackage{multirow}
\usepackage{scalefnt}
\usepackage{enumerate}
\usepackage[printonlyused]{acronym}
\usepackage{centernot}
\usepackage{caption}
\usepackage{color}

\theoremstyle{remark} 
\theoremstyle{remark}

\theoremstyle{remark} 
\newtheorem*{remarkA}{Remark}

\newcommand{\mC}{\mathcal{C}}

\newcommand{\yvec}{\mathbf{y}}
\newcommand{\xvec}{\mathbf{x}}
\newcommand{\cvec}{\mathbf{c}}
\newcommand{\zvec}{\mathbf{z}}

\newcommand{\nvec}{\mathbf{n}}

\newcommand{\bvec}{\mathbf{b}}

\newcommand{\vvec}{\mathbf{v}}

\newcommand{\mvec}{\mathbf{m}}

\newcommand{\dsM}{\mathds{M}}

\newcommand{\dsI}{\mathds{I}}

\newcommand{\dsG}{\mathds{G}}

\long\def\symbolfootnote[#1]#2{\begingroup\def\thefootnote{\fnsymbol{footnote}}\footnote[#1]{#2}\endgroup}

\newcommand{\ConstraintLen}{q}

 {\begin{list}{}%
         {\setlength{\leftmargin}{#1}}%
         \item[]%
 }
 {\end{list}}

\title{A New Approach to UEP-HARQ via \\ Convolutional Codes}

\author{
\IEEEauthorblockN{Annabel Sharon Shitrit, Yonathan Murin, {\em Member, IEEE}, Ron Dabora, {\em Senior Member, IEEE}, and Osnat Keren}\thanks{
A. Sharon Shitrit and R. Dabora (Email: ron@ee.bgu.ac.il) are with the Department of Electrical and Computer Engineering, Ben-Gurion University, Beer-Sheva 84105, Israel. Y. Murin (Email: moriny@stanford.edu) is with the
Department of Electrical Engineering, Stanford University, Stanford, CA 94305, USA. O. Keren (Email: Osnat.Keren@biu.ac.il) is with the Faculty of Engineering, Bar-Ilan University, Ramat-Gan 52900, Israel.
This work was supported by the Israeli Science Foundation under Grant 396/11.
}\vspace{-1cm}}

\begin{document}

\maketitle

\thispagestyle{empty} 
\pagestyle{empty}

\begin{abstract}
	This paper presents a novel type-II hybrid automatic repeat request (HARQ) transmission scheme which is based on pruned convolutional codes (CCs) and supports unequal error protection (UEP).
The data to be transmitted is assumed to consist of important bits (IB) and standard bits (SB).
In the proposed scheme, all the bits are first encoded using a single mother CC and transmitted over the channel. If a decoding error is detected in the IB, then retransmission of {\em only} the IB takes place using a CC, typically with a better error correction capability.
Next, taking advantage of the properties of pruned CCs, the decoded IB are used to increase the decoding reliability for the SB.
Numerical simulations indicate that the proposed scheme offers strong protection for the IB along with improved reliability for the remaining bits compared to other UEP schemes.
\end{abstract}


\section{Introduction}

Hybrid automatic repeat request (HARQ) schemes are frequently employed in modern wireless transmission protocols to counter the effects of channel variations on the system performance. These schemes combine forward error correction coding with retransmissions, thus increasing the overall throughput.
While traditionally all the data bits are treated as having equal importance, there are many modern practical applications which require different protection levels for different parts of the information. Two common examples are video streaming applications \cite{Mohr-Riskin:2000} and scalable image representation \cite{Salemi-Desset:2008}. To apply different protection levels, coding schemes which provide unequal error protection (UEP) are employed.
In this work we propose a novel HARQ scheme which employs UEP and is based on pruned convolutional codes (CCs). A description of pruned CCs can be found in \cite{nested_conv_2} and \cite{UEP_2002_prunning}.

The HARQ scheme developed in this work is of type-II, commonly referred to as incremental redundancy (IR)-HARQ. Differently from type-I HARQ schemes, in which the entire encoded data block is retransmitted until it is correctly decoded (or until the maximum number of retransmissions is reached), in IR-HARQ each retransmission carries incremental redundant information about the data block.
Furthermore, while in type-I HARQ blocks, which are not successfully decoded, are discarded, in IR-HARQ decoding is based on both previously received blocks and the currently received block.
When combined with CC, IR-HARQ is commonly implemented via puncturing \cite{RCPC}, namely, only a subset of the codeword symbols is transmitted, and when a decoding error is detected, only the punctured bits are transmitted.
IR-HARQ schemes based on punctured CCs were first presented in \cite{RCPC}.
The work \cite{RCPC} presented the class of rate-compatible punctured CCs (RCPCCs) which are constructed from a rate $1/N$ mother CC, where $N$ is a positive integer, and developed an IR-HARQ scheme based on these codes. Later, \cite{Kallel_1990} and \cite{Kallel_1995} extended the approach of \cite{RCPC} and presented IR-HARQ schemes based on punctured CCs and code combining.

	The schemes presented in \cite{RCPC}--\cite{Kallel_1995} implement IR-HARQ via CCs while {\em equally protecting} all the transmitted information. A different line of works studied UEP via CCs without HARQ \cite{nested_conv_2}, \cite{UEP_2002_prunning}, \cite{UEP_2008_prunning}, \cite{{Hanzo_2013}} and \cite{UEP_CC_2014}. The work \cite{nested_conv_2} studied the problem of concatenating inner and outer CCs, focusing on the mechanism for partitioning the inner CC into several subcodes with different error correction capabilities. The partitioning method presented in \cite{nested_conv_2} is based on subcodes constructed via removing specific paths in the code's trellis, along with an algorithm for finding the partitioning which maximizes the free distance of the subcodes. We refer to this approach as {\em path pruning} or {\em pruned CCs}. 	
The work \cite{UEP_2002_prunning} also considered path pruning, and proposed a cascaded pruning procedure which maintains the continuity of the encoding path and avoids unpredictable path distance decrease.
The work \cite{UEP_2008_prunning} extended this approach by combining path pruning and puncturing.
Recently, UEP via CCs was proposed in \cite{Hanzo_2013} for the wireless transmission of video, which consists of a base layer and enhancement layers. The scheme of \cite{Hanzo_2013} jointly encodes all the video layers and uses the decoded enhancement layers to assist in decoding the base layer. This is in contrast to our proposed scheme  in which decoding the important data assists in decoding the rest of the data.
Finally, \cite{UEP_CC_2014} proposed a transmission scheme which uses UEP Raptor codes at the application layer and UEP RCPCCs at the physical layer for prioritized video packets.


Several works studied coding for UEP-HARQ.
The works \cite{SJKE:2002} and \cite{QJSZY:2008} proposed UEP-HARQ via RCPCCs and turbo codes, respectively.
The work \cite{UEP_2015} studied the transmission of layered video and proposed an adaptive truncated HARQ scheme which is based on CCs with separate encoding of each layer via a different RCPCC, and with a type-I HARQ with maximal ratio combining.
Note that in \cite{SJKE:2002}--\cite{UEP_2015}, data of different importance levels are encoded {\em separately} and parallel HARQ schemes are used.
On the other hand, in our scheme all the data is {\em jointly} encoded, and retransmission of the important data also improves the reliability of the standard data.

%

 {\textbf {\slshape Main contributions}:} We present a novel UEP-HARQ scheme which is based on pruned CCs. The data stream is assumed to consist of important bits (IB) and standard bits (SB), where reliable decoding of the IB is required, while residual bit error rate (BER) in the SB is acceptable.
 At the first transmission, the entire sequence of bits is encoded via a single CC. In case of decoding failure of the IB, retransmission is applied {\em only} to these bits, thus, implementing UEP. \textcolor{black}{By using a special arrangement of the data bit sequence, the proposed scheme can use the decoded IB to minimize the BER for the SB. To the best of our knowledge, this is the first scheme which uses the retransmission of the IB to improve the BER in the SB.}
Compared to equal error protection (EEP) schemes such as the scheme of \cite{Kallel_1990}, the proposed scheme achieves higher reliability of the IB, at the cost of higher BER in the SB, thereby offering a tradeoff between the levels of protection.
Compared to schemes which implement UEP via separate encoding, and use the retransmissions for sending only the IB, the proposed scheme achieves lower BER in the SB for the same level of protection of the IB.

%

The rest of this paper is organized as follows: Section \ref{sec:probDef} presents the problem formulation and briefly recalls pruned CCs, Section \ref{sec:uep-harq_scheme} describes the novel UEP-HARQ scheme, Section \ref{sec:simulations} details the numerical simulations, and Section \ref{sec:conc} concludes the paper.

{\textbf{\slshape Notation}:} \textcolor{black}{Boldface lowercase letters are used to denote row vectors, e.g., $\xvec$, where $|\xvec|$ denotes the length of $\xvec$. Double-stroke capital letters are used to denote matrices, e.g., $\dsM$, where $\text{det}(\dsM)$ is used to denote the determinant of the matrix $\dsM$ and $\dsI_K$ denotes the identity matrix of size $K \mspace{-3mu} \times \mspace{-3mu} K$.
Calligraphic capital letters are used to denote codebooks, e.g., $\mC$, where $\mC(n,k,d,q)$ denotes a CC with $n$ output bits per each $k$-tuple input bits, a free distance $d$, and a constraint~length~$q$.}

\section{Problem Formulation and Preliminaries} \label{sec:probDef}

\subsection{Problem Formulation} \label{subsec:probForm}

We consider the transmission of a bit-sequence $\mvec$, over a wireless channel with noiseless 1-bit feedback.
The sequence $\mvec$ consists of IB $\mvec_{1}$ and SB $\mvec_{2}$.
The objective of the transmitter is to reliably convey $\mvec_{1}$ to the receiver, while minimizing the BER for $\mvec_{2}$. In order to achieve improved reliability for $\mvec_{1}$, it is first encoded with a cyclic redundancy check (CRC) code; the CRC bits are denoted by $\mvec_{3}$.
The sequences $\{\mvec_{i}\}_{i = 1}^3$ are combined into a sequence $\bvec$ which is encoded via a CC into a codeword $\cvec$ of length $M$.
The codeword $\cvec$ is then BPSK modulated, resulting in the sequence $\xvec$ which is transmitted over a block-fading channel. The elements of $\xvec$ are assumed to be zero-mean with unit variance, and the channel output is given by:
\begin{equation}
	\yvec = h \cdot \xvec + \zvec, \quad \zvec \sim \mathcal{N}(\boldsymbol{0}, \dsI_M), \label{eq:chan_model}
\end{equation}

\noindent where $h$ is the channel gain, which randomly varies between transmission blocks, and is known only at the receiver. The channel signal-to-noise ratio is therefore given by $\text{SNR}_{\text{c}} \mspace{-3mu} = \mspace{-3mu} E\{|h|^2\}$, where $E\{\cdot\}$ denotes the stochastic expectation.
The receiver decodes $\mvec_1$, $\mvec_2$ and $\mvec_3$ from $\yvec$, and the CRC $\mvec_3$ is used to verify that the IB $\mvec_1$ are correctly decoded. If $\mvec_1$ was decoded correctly, then the receiver sends an ACK to the transmitter; If decoding of $\mvec_1$ fails, the receiver sends a NACK, and a retransmission takes place. Retransmissions are repeated until $\mvec_1$ is decoded correctly or until a predefined number of maximal retransmissions is~reached.

\subsection{A Short Review of Pruned CCs} \label{subsec:pruned_CCs}
Let $\mC_{\text{A}}(n \mspace{-3mu} = \mspace{-3mu} N, k \mspace{-3mu} = \mspace{-3mu} 1, d_{\text{A}}, q_{\text{A}})$ be a binary CC. The work \cite{nested_conv_2} proposed a procedure for constructing $q_{\text{A}}$ subcodes $\{\mC_{\text{P},j}\}_{j=0}^{q_{\text{A}}-1}$ from $\mC_{\text{A}}$, with improved distance properties.
We refer to the subcodes $\{\mC_{\text{P},j}\}_{j=0}^{q_{\text{A}}-1}$ as pruned CCs and in this section we briefly review their construction. An extension to any integer $k$ and $n$ can be found in \cite{UEP_2002_prunning}.
Let the encoder input $\bvec$ be a binary sequence of length $L \mspace{-3mu} \cdot \mspace{-3mu} \ConstraintLen_\text{A}, L \mspace{-3mu} = \mspace{-3mu}  \lceil ({\textstyle \sum_{i=1}^3{|\mvec_i|} }) /  \ConstraintLen_\text{A} \rceil$.
The sequence $\bvec$ can be represented as a set of $L$ sequences $\{\bvec_t\}_{t=1}^L$, each of length $\ConstraintLen_\text{A}$.
As was shown in \cite{UM_code}, $\mC_{\text{A}}$ can be represented as a unit memory (UM) code $\mC_{\text{A}}^{\text{UM}}$ $(n \mspace{-3mu} = \mspace{-3mu} N \mspace{-1mu} \cdot \mspace{-1mu} \ConstraintLen_{\text{A}}, k \mspace{-3mu} = \mspace{-3mu} \ConstraintLen_\text{A}, d_{\text{A}}, 1)$, with the encoding rule:
\begin{align}
	\cvec_{t} = \bvec_{t} \cdot \dsG_0 + \bvec_{t-1} \cdot \dsG_1, \label{UMcode}
\end{align}

\noindent where $\dsG_0$ and $\dsG_1$ are binary matrices of size $\ConstraintLen_\text{A} \mspace{-2mu} \times N \mspace{-7mu} \cdot \mspace{-3mu} \ConstraintLen_{\text{A}}$, and $\{ \cvec_{t} \}_{t=1}^L$ are output vectors, each of length $N \mspace{-1mu} \cdot \mspace{-1mu} \ConstraintLen_{\text{A}}$. In \cite{nested_conv_2} the $j$'th subcode, denoted by $\mC_{\text{P},j}$, is constructed by setting the first $j$ rows of $\dsG_0$ and of $\dsG_1$ to zero, which implies that $\mC_{\text{P},\ConstraintLen_\text{A}-1} \mspace{-3mu} \subseteq \mspace{-3mu} \mC_{\text{P},\ConstraintLen_\text{A}-2} \mspace{-3mu} \subseteq \mspace{-3mu} \cdots \mspace{-3mu} \subseteq \mspace{-3mu} \mC_{\text{P},0} \mspace{-3mu} = \mspace{-3mu} \mC_{\text{A}}^{\text{UM}}$. Therefore, letting $d_j$ denote the free distance of $\mC_{\text{P},j}$, we have $d_{\ConstraintLen_\text{A}-1} \mspace{-3mu} \ge \mspace{-3mu} d_{\ConstraintLen_\text{A}-2} \mspace{-3mu} \ge \mspace{-3mu} \dots \mspace{-3mu} \ge \mspace{-3mu} d_{0} \mspace{-3mu} = \mspace{-3mu} d_{\text{A}}$.
We note that $\mC_{\text{P},j}$ can be equivalently constructed by setting the first $j$ bits in each $\bvec_t$ to zero, resulting in a code whose rate is $\frac{\ConstraintLen_\text{A}-j}{N \mspace{-1mu} \cdot \mspace{-1mu} \ConstraintLen_{\text{A}}}$. This observation implies that $\mC_{\text{P},j}$ can be equivalently obtained by pruning specific paths from the trellis of $\mC_{\text{A}}^{\text{UM}}$.

It was observed in \cite[Sec. 2]{nested_conv_2} that the matrices $\dsG_0$ and $\dsG_1$ are not unique, namely, the code $\mC_{\text{A}}$ can also be obtained by using the matrices $\tilde{\dsG}_i \mspace{-3mu} = \mspace{-3mu} \dsM \mspace{-3mu} \cdot \mspace{-3mu} \dsG_i , i \mspace{-3mu} = \mspace{-3mu}0,1$, where $\dsM$ is a binary scrambler matrix of size $\ConstraintLen_\text{A} \mspace{-4mu} \times \mspace{-3mu} \ConstraintLen_\text{A}$ with $\text{det}(\dsM) \mspace{-4mu} = \mspace{-3mu} 1$ over the binary field.
In fact, \cite{nested_conv_2} showed that different encoding matrices can result in different $\{d_j\}_{j= 0}^{\ConstraintLen_\text{A}-1}$, and proposed an algorithm for finding the matrix $\dsM$ that maximizes $\{d_j\}_{j= 0}^{\ConstraintLen_\text{A}-1}$ \cite[pg. 487]{nested_conv_2}.

\section{The Main Result: A New UEP-HARQ Scheme} \label{sec:uep-harq_scheme}

In this section we derive our novel UEP-HARQ scheme.
In Subsection \ref{subsec:HARQ_scheme} we describe the main idea, then, in Subsection \ref{subsec:mapping_and_v} we show how to implement the proposed approach using pruned CCs, and finally in Subsection \ref{subsec:proposed_scheme} we provide a detailed description of the overall scheme. To illustrate the concept we base the explanations on hard decoding, though the proposed scheme straight-forwardly applies to soft decoding.

\subsection{Projection on a Subcode} \label{subsec:HARQ_scheme}
In the proposed scheme the IR transmission is designed to {\em simultaneously}: 1) Convey additional information on the IB, thereby increasing their decoding reliability; and, 2) Project the channel output of the first transmission onto a subcode with a larger free distance in order to increase the decoding reliability of the SB.
Since the IR transmission affects differently the IB and SB, the proposed scheme provides UEP.

As discussed in Section \ref{subsec:probForm}, the IB, $\mvec_1$, are protected by a CRC code whose parity-check bits are represented by $\mvec_3$.
The combined sequence $\bvec$ is encoded into a codeword $\cvec_\text{A} \mspace{-3mu} \in \mspace{-3mu} \mC_{\text{A}}^{\text{UM}} \mspace{-3mu} \equiv \mspace{-3mu} \mC_{\text{A}}$
and transmitted over the channel. Note that in this initial transmission $\mvec_1$ and $\mvec_2$ are equally protected.
If decoding of $\mvec_1$ fails, a NACK message is sent to the transmitter. Upon receiving a NACK, the transmitter projects $\cvec_{\text{A}}$ onto a subcode of $\mC_{\text{A}}^{\text{UM}}$, which is one of its pruned CCs, described in Section \ref{subsec:pruned_CCs}.
Let this subcode be $\mC_{\text{P},j}$.
The projection onto $\mC_{\text{P},j}$ is done via mapping the codeword $\cvec_{\text{A}}$ to a specific $\cvec_{\text{P},j}(\cvec_{\text{A}}) \in \mC_{\text{P},j}$ (the mapping is detailed in Section \ref{subsec:mapping_and_v}, and in Section \ref{subsec:proposed_scheme} the subcode selection, i.e., choosing the value of $j$, is described). As $\mC_{\text{P},j} \mspace{-2mu} \subseteq \mspace{-2mu}  \mC_{\text{A}}^{\text{UM}}$, for each $\cvec_{\text{A}} \in \mC_{\text{A}}^{\text{UM}}$ we can write:
\begin{align}
	\cvec_{\text{A}} = \cvec_{\text{P},j}(\cvec_{\text{A}}) + \vvec(\cvec_{\text{A}}, \cvec_{\text{P},j}), \quad \vvec(\cvec_{\text{A}}, \cvec_{\text{P},j}) \in \mC_{\text{A}}^{\text{UM}}, \label{eq_bias}
\end{align}

\noindent where $\vvec(\cvec_{\text{A}}, \cvec_{\text{P},j})$ is referred to as the projection offset.
By sending $\vvec(\cvec_{\text{A}}, \cvec_{\text{P},j})$ to the receiver as IR, the receiver can project the channel output observed at the first transmission onto the pruned CC $\mC_{\text{P},j}$.
Let $\tilde{\cvec}_{\text{A}} \mspace{-3mu} = \mspace{-3mu} \cvec_{\text{A}} \mspace{-2mu} + \mspace{-2mu} \nvec_{\text{A}}$ be a binary sequence corresponding to the received channel output at the first transmission, $\yvec$, where $\nvec_{\text{A}}$ represents the errors due to channel noise, and let $\hat{\vvec}(\cvec_{\text{A}}, \cvec_{\text{P},j})$ be the decoded version of $\vvec(\cvec_{\text{A}}, \cvec_{\text{P},j})$ after retransmission. Then, assuming that  $\vvec(\cvec_{\text{A}}, \cvec_{\text{P},j}) = \hat{\vvec}(\cvec_{\text{A}}, \cvec_{\text{P},j})$, the receiver calculates:
\begin{align}
	 \mspace{-6mu} \tilde{\cvec}_{\text{A}} \mspace{-3mu} - \mspace{-3mu} \hat{\vvec}(\cvec_{\text{A}}, \cvec_{\text{P},j}) \mspace{-2mu} = \mspace{-2mu} \cvec_{\text{A}} \mspace{-3mu} + \mspace{-3mu} \nvec_{\text{A}} - \mspace{-3mu} \vvec(\cvec_{\text{A}}, \cvec_{\text{P},j}) \mspace{-2mu} = \mspace{-2mu} \cvec_{\text{P},j}(\cvec_{\text{A}}) \mspace{-3mu} + \mspace{-3mu} \nvec_{\text{A}}. \label{eq:dec_cp}
\end{align}

\noindent Thus, the receiver obtains $\cvec_{\text{P},j}(\cvec_{\text{A}})$ corrupted by the noise vector~$\nvec_{\text{A}}$. In the following subsections we show that by properly constructing the sequence $\bvec$, the IB can be transmitted as IR.
Furthermore, with the proposed construction of $\bvec$, after projection at the receiver via \eqref{eq:dec_cp}, the SB are protected by $\mC_{\text{P},j}$ whose error correction capability is better than that of $\mC_{\text{A}}^{\text{UM}}$.
We further show that with the proposed $\cvec_{\text{P},j}(\cvec_{\text{A}})$ the number of possible $\vvec(\cvec_{\text{A}}, \cvec_{\text{P}_\text{j}})$ sequences is significantly lower than $2^{|\vvec(\cvec_{\text{A}}, \cvec_{\text{P},j})|}$. Thus, instead of directly sending $\vvec(\cvec_{\text{A}}, \cvec_{\text{P},j})$ over the channel as IR, we send a shorter sequence, denoted by $\vvec_0(\cvec_{\text{A}}, \cvec_{\text{P},j})$, which can be protected by a CC with a significantly better error correction capability compared to $\mC_{\text{A}}^{\text{UM}}$. This supports the assumption $\hat{\vvec}(\cvec_{\text{A}}, \cvec_{\text{P},j}) = \vvec(\cvec_{\text{A}}, \cvec_{\text{P},j})$.

\subsection{Calculating $\cvec_{\text{P},j}(\cvec_{\text{A}})$ and the Projection Offset $\vvec(\cvec_{\text{A}}, \cvec_{\text{P},j})$} \label{subsec:mapping_and_v}

We now describe the calculation of $\cvec_{\text{P},j}(\cvec_{\text{A}}) \mspace{-3mu} \in \mspace{-3mu} \mC_{\text{P},j}(N \mspace{-2mu} \cdot \mspace{-2mu} \ConstraintLen_{\text{A}}, \mspace{-2mu} \ConstraintLen_{\text{A}} \mspace{-2mu} - \mspace{-2mu} j \mspace{-2mu},\mspace{-2mu} d_j, \mspace{-2mu} 1)$ and $\vvec(\cvec_{\text{A}}, \cvec_{\text{P},j})$. Recall that the length of $\bvec$ is $L \mspace{-2mu} \cdot \mspace{-2mu} \ConstraintLen_{\text{A}} $, otherwise, append (at most) $\ConstraintLen_{\text{A}}\mspace{-4mu} -\mspace{-4mu} 1$ zeros to $\bvec$ to satisfy this requirement; as $\ConstraintLen_{\text{A}}\mspace{-4mu} \ll \mspace{-4mu} |\bvec|$ the impact of this padding on the throughput is negligible.
Let $f(\{\bvec_t\}_{t=1}^L)$ denote the extension of the UM encoding operation stated in \eqref{UMcode} to the entire sequence $\bvec$ in the obvious manner.
The codeword $\cvec_{\text{A}}$ is mapped into $\cvec_{\text{P},j}(\cvec_{\text{A}})$ by: 1) Setting the first $j$ bits in each $\bvec_t$ to zero, resulting in $\bvec_{t,j}$, see Section \ref{subsec:pruned_CCs}; and, 2) Applying the encoding $\cvec_{\text{P},j}(\cvec_{\text{A}}) \mspace{-3mu} = \mspace{-3mu} f(\{\bvec_{t,j}\}_{t=1}^L)$. From Eq. \eqref{eq_bias} we~obtain:
\begin{align}
	\vvec(\cvec_{\text{A}}, \cvec_{\text{P},j}) & = \cvec_{\text{A}} - \cvec_{\text{P},j}(\cvec_{\text{A}}) \nonumber \\
	& \stackrel{(a)}{=} f \left( \{\bvec_t\}_{t=1}^L - \{\bvec_{t,j}\}_{t=1}^L \right) \nonumber \\
	%
	& \stackrel{(b)}{=} f \left( \{ b_{t,1}, b_{t,2}, \dots, b_{t,j}, 0, \dots , 0 \}_{t=1}^L \right), \label{eq:v_cc_explicit}
\end{align}

\noindent where (a) follows from the linearity of CCs, and (b) follows from the definition of $\bvec_{t,j}$. Eq. \eqref{eq:v_cc_explicit} implies that $\vvec(\cvec_{\text{A}}, \cvec_{\text{P},j})$ is obtained by encoding only the information bits encoded in $\cvec_{\text{A}}$ but not in $\cvec_{\text{P},j}(\cvec_{\text{A}})$. Thus, $\vvec(\cvec_{\text{A}}, \cvec_{\text{P},j})$ contains the information that is not present in $\cvec_{\text{P},j}(\cvec_{\text{A}})$.
Clearly, the zeros in \eqref{eq:v_cc_explicit} need not be transmitted, thus, $\vvec(\cvec_{\text{A}}, \cvec_{\text{P},j})$ is uniquely represented~by:
\begin{align}
	\vvec_0(\cvec_{\text{A}}, \cvec_{\text{P},j}) =  \{ b_{t,1}, b_{t,2}, \dots, b_{t,j} \}_{t=1}^L. \label{eq:v0_compact}
\end{align}

\noindent The transmission of $\vvec_0(\cvec_{\text{A}}, \cvec_{\text{P},j})$ over the channel is described in Step 4) in Section \ref{subsec:proposed_scheme}.
After decoding $\vvec_0(\cvec_{\text{A}}, \cvec_{\text{P},j})$, the receiver appends $\ConstraintLen_{\text{A}} \mspace{-3mu} - \mspace{-3mu} j \mspace{-2mu}$ zeros to each $\{ \hat{b}_{t,1}, \hat{b}_{t,2}, \dots, \hat{b}_{t,j} \}$, and encodes the resulting sequence via $f(\cdot)$ to obtain $\hat{\vvec}(\cvec_{\text{A}}, \cvec_{\text{P},j})$, which is applied to obtain $\hat{\cvec}_{\text{P},j}(\cvec_{\text{A}})$ via \eqref{eq:dec_cp}.
Next, we describe in detail the overall proposed scheme.

\subsection{Detailed Description of the UEP-HARQ Scheme} \label{subsec:proposed_scheme}

Our scheme uses a mother CC $\mC_{\text{A}}^{\text{UM}}(N \mspace{-1mu} \cdot \mspace{-1mu} \ConstraintLen_{\text{A}}, \ConstraintLen_{\text{A}}, d_\text{A},1)$.
In the proposed scheme, upon receiving a NACK, the transmitter sends $\mvec_1$ and $\mvec_3$ via $\vvec_0(\cvec_{\text{A}}, \cvec_{\text{P},j})$. Then, $\vvec(\cvec_{\text{A}}, \cvec_{\text{P},j})$ is used by the receiver to improve the reliability of $\mvec_2$.
Assume that each transmitted block consists of $M$ channel uses.
The proposed scheme consists of the following steps:

{\em 1) Choosing the subcode $\mC_{\textrm{P},j}$:}
As $\vvec_0(\cvec_{\text{A}}, \cvec_{\text{P},j})$ should support the transmission of $\mvec_1$ and $\mvec_3$, from Eq. \eqref{eq:v0_compact} it follows that $j$ should satisfy $|\mvec_1| \mspace{-2mu} + \mspace{-2mu} |\mvec_3| \le |\vvec_0(\cvec_{\text{A}}, \cvec_{\text{P},j})| = j\cdot L$.
\noindent On the other hand, minimizing $|\vvec_0(\cvec_{\text{A}}, \cvec_{\text{P},j})|$ enables using stronger CC at the retransmission phase. Therefore, we propose to choose $\mC_{\textrm{P},j}$ such that $j =  \left\lceil (|\mvec_1| + |\mvec_3|)/L \right\rceil$.
%

{\em 2) Constructing the sequence $\bvec$:}
Following \eqref{eq:v_cc_explicit}, the bits $\{ b_{t,1}, b_{t,2}, \dots, b_{t,j} \}_{t=1}^L$ are allocated to $\mvec_1$ and $\mvec_3$, while the rest of the bits of $\bvec$ are allocated to $\mvec_2$.

{\em 3) The first transmission:}
The transmitter encodes $\bvec$ using $\mC_{\text{A}}^{\text{UM}}$, modulates the encoded vector, and transmits it over the channel. The receiver decodes the vector $\bvec$ using the channel output $\yvec$, and checks the correctness of the decoded $\mvec_1$ using the decoded CRC $\mvec_3$. If $\mvec_1$ was correctly decoded, the receiver sends an ACK to the transmitter and the algorithm terminates. Otherwise, the receiver sends a NACK.

{\em 4) Retransmission:}
Upon receiving a NACK, the transmitter encodes the sequence $\vvec_0(\cvec_{\text{A}}, \cvec_{\text{P},j})$ using a CC, denoted by $\mC_{\text{IR}}$, and sends it over the channel. $\mC_{\text{IR}}$ is selected to have a constraint length $\ConstraintLen_\text{A}$ in order to obtain roughly the same decoding complexity as in the first transmission. For example, if {\em retransmission} spans $M$ channel uses and conveys only the $L$ IB, then the rate of $\mC_{\text{IR}}$ is:
\begin{equation}
	R_{\text{IR}} = \frac{j \cdot L}{M} = \frac{j \cdot L}{\left( L \mspace{-3mu} \cdot \mspace{-3mu} \ConstraintLen_\text{A} \mspace{-2mu} + \mspace{-2mu} \ConstraintLen_\text{A} \right)N}. \label{eq:R_B}
\end{equation}

\noindent In \eqref{eq:R_B} $M$ can be obtained from analyzing the first transmission: Recall that $|\bvec| \mspace{-3mu} = \mspace{-3mu} L \mspace{-3mu} \cdot \mspace{-3mu} \ConstraintLen_\text{A}$, and note that $\ConstraintLen_\text{A}$ tail bits are appended to $\bvec$ for termination. Recalling that the rate of $\mC_{\text{A}}$ is $1/N$ we obtain $M \mspace{-3mu} = \mspace{-3mu} (L \mspace{-3mu} + \mspace{-3mu} 1)\ConstraintLen_\text{A} N$.
We emphasize that any code rate can be used for retransmission, as long as the IB are sent during the retransmission phase.

{\em 5) Decoding the retransmission at the receiver:} The receiver decodes $\vvec_0(\cvec_{\text{A}}, \cvec_{\text{P},j})$ (and therefore $\mvec_1$ and $\mvec_3$) and maps it to $\vvec(\cvec_{\text{A}}, \cvec_{\text{P},j})$ as described in \textcolor{black}{Section \ref{subsec:mapping_and_v}}. It then computes \eqref{eq:dec_cp} and decodes $\mvec_2$ based on the code $\mC_{\text{P},j}$; this is done using a Viterbi decoder which accounts for the pruned paths in the trellis of $\mC_{\text{A}}^{\text{UM}}$ \textcolor{black}{\cite{nested_conv_2}}. If $\mvec_1$ passes the CRC test, the scheme is terminated, otherwise, $\vvec_0(\cvec_{\text{A}}, \cvec_{\text{P},j})$ is retransmitted as detailed in Step 4), where in Step 5), $\vvec_0(\cvec_{\text{A}}, \cvec_{\text{P},j})$ is decoded via code combining, see e.g., \textcolor{black}{\cite[Sec. II]{Kallel_1990}}.

\begin{remarkA}[Soft decoding]
The proposed scheme also supports soft decoding. Applying soft decoding for the codes $\mC_{\text{A}}^{\text{UM}}, \mC_{\text{IR}}$ and $\mC_{\text{P},j}$ is straightforward.
To implement the "soft" calculation of \eqref{eq:dec_cp} we note that when BPSK modulation is used, then the subtraction (XOR) of $\hat{\vvec}(\cvec_{\text{A}}, \cvec_{\text{P},j})$ in \eqref{eq:dec_cp} can be implemented via element-wise product of the BPSK representation of $\hat{\vvec}(\cvec_{\text{A}}, \cvec_{\text{P},j})$, and the channel output of the first transmission ${\bf y}$, see \cite[Eq.~(11)]{soft_decoding}.
%
%

\end{remarkA}

\section{Numerical Simulations} \label{sec:simulations}

\subsection{Simulation Setup}

\begin{figure*}[t]
	\normalsize
	\captionsetup{font=small}
	\centering
	\begin{minipage}{.48\textwidth}
		\centering
		\includegraphics[width=1\linewidth]{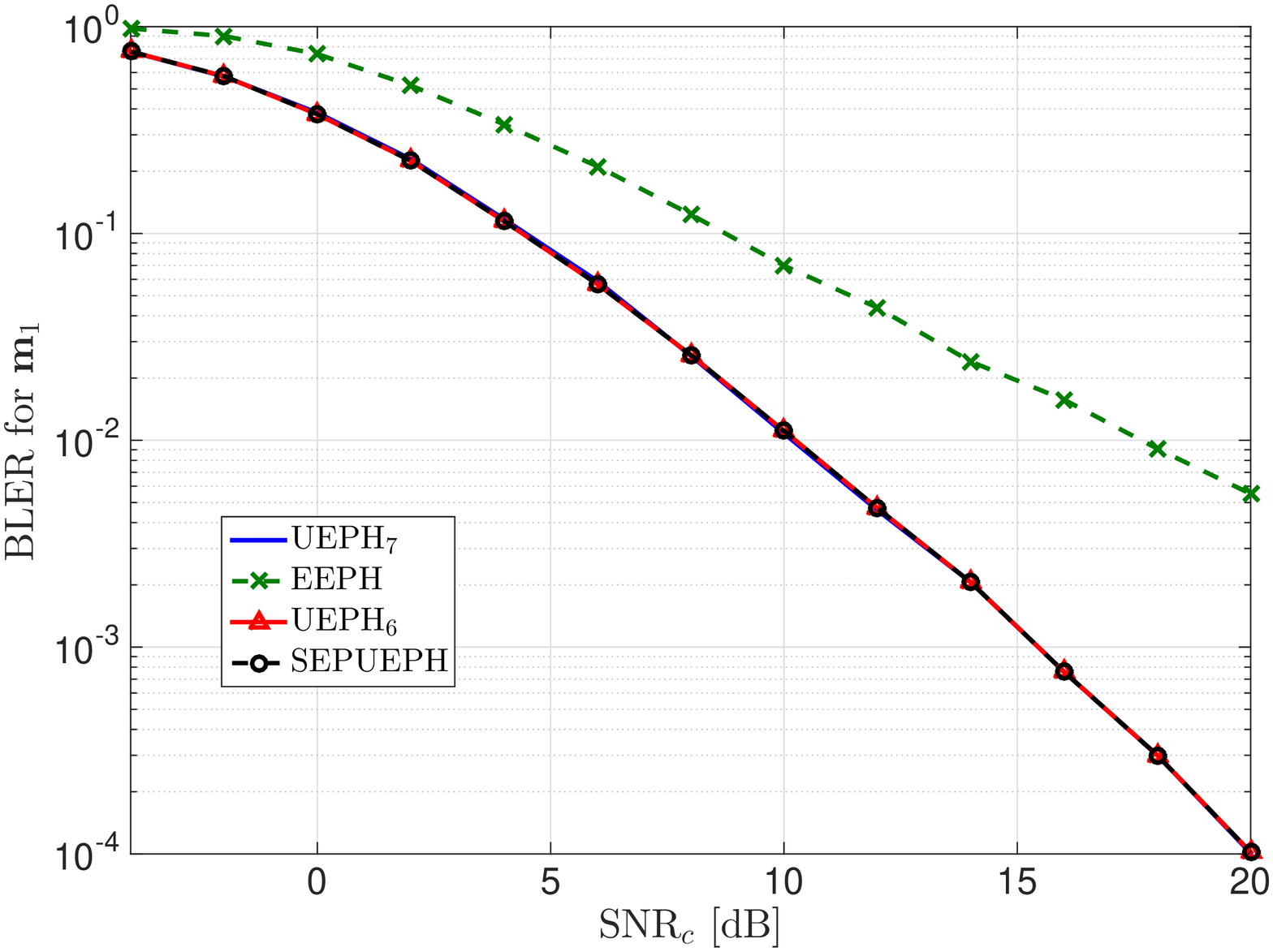}
		\vspace{-0.15cm}
		\captionof{figure}{\scriptsize BLER in decoding $\mvec_1$ vs. $\text{SNR}_{\text{c}}$, for block Rayleigh fading channel.}
		\label{fig:BLER}
		\vspace{-0.3cm}
	\end{minipage}
	\hspace{0.4cm}
	\begin{minipage}{.48\textwidth}
		\centering
		\includegraphics[width=1\linewidth]{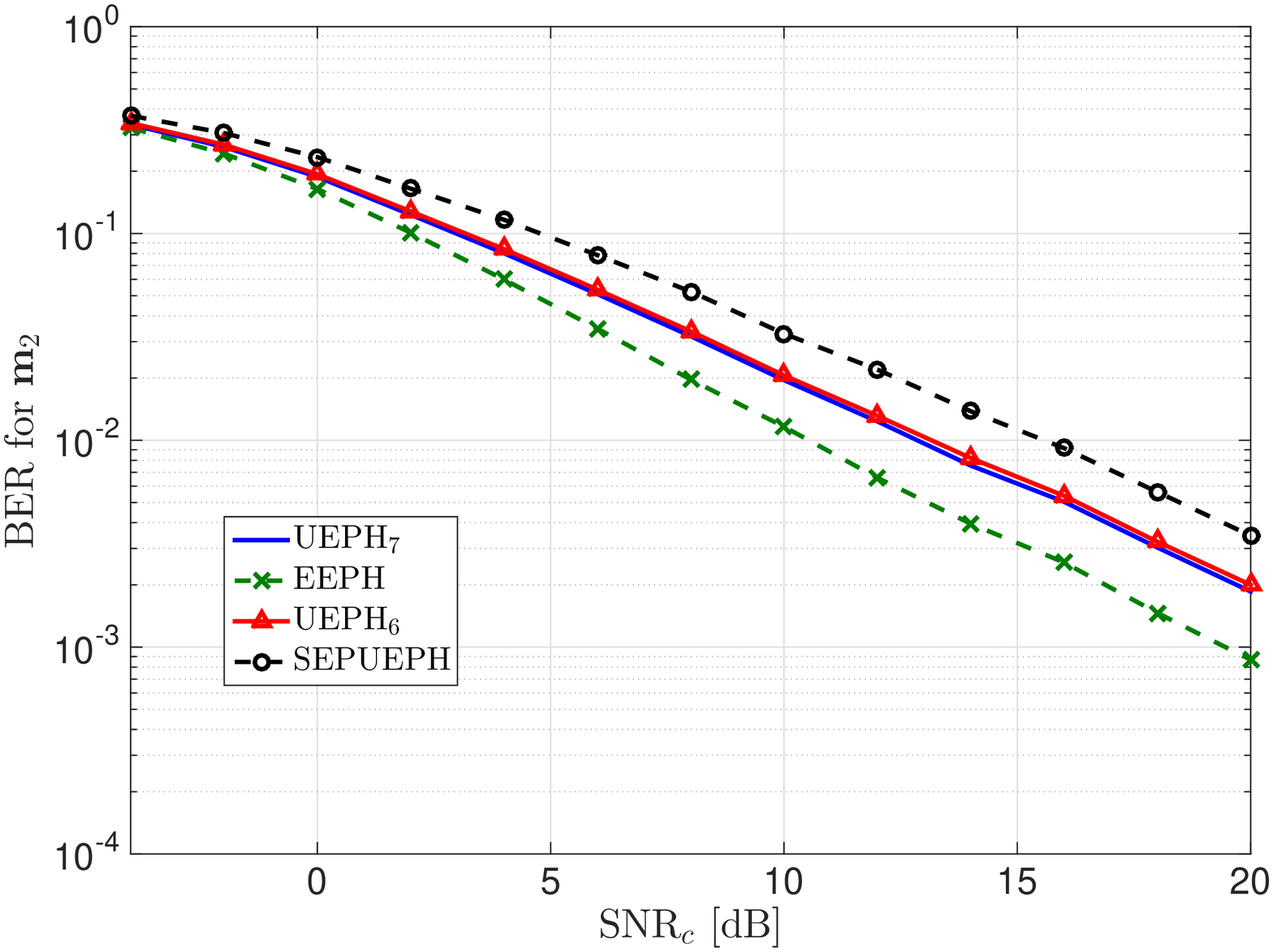}
		\vspace{-0.15cm}
		\captionof{figure}{\scriptsize BER in decoding $\mvec_2$ vs. $\text{SNR}_{\text{c}}$, for block Rayleigh fading channel.}
		\label{fig:BER}
		\vspace{-0.3cm}
	\end{minipage}
\end{figure*}

In the simulations we set $\mC_{\text{A}}$ to be a rate $1/2$ CC with generator polynomials in octal form $G_1 \mspace{-3mu} = \mspace{-3mu} 15$ and $G_2 \mspace{-3mu} = \mspace{-3mu} 17$, \textcolor{black}{which implies that $\ConstraintLen_\text{A} \mspace{-4mu} = \mspace{-4mu} 3$ and $d_{\text{A}}\mspace{-4mu}=\mspace{-4mu}6$}.
The bit sequence $\bvec$ consists of 1500 bits, we assume ideal CRC and set $|\mvec_3| \mspace{-3mu} = \mspace{-3mu} 0$, and also let $|\mvec_1| \mspace{-3mu} = \mspace{-3mu} 500$ and $|\mvec_2| \mspace{-3mu} = \mspace{-3mu} 1000$. Thus, $L \mspace{-3mu} = \mspace{-3mu} 500$ and $M \mspace{-3mu} = \mspace{-3mu} 3006$.
UM encoding with the optimal encoding matrices specified in \cite[pg. 486]{nested_conv_2} is implemented. From Section \ref{subsec:proposed_scheme}1 we obtain $j \mspace{-3mu} = \mspace{-3mu} 1$ which leads to $d_1 \mspace{-3mu} = \mspace{-3mu} 7$.
\textcolor{black}{For the retransmission we use \eqref{eq:R_B} to obtain $R_{\text{IR}} = 500/3006$. The sequence to be retransmitted is obtained by appending $\ConstraintLen_{\text{IR}} \mspace{-6mu} = \mspace{-6mu} \ConstraintLen_{A} \mspace{-6mu} = \mspace{-6mu} 3$ tail bits to $\mvec_1$ which will be encoded via a CC of rate $1/6$ with generator polynomials $\{G_1, G_2, G_3, G_1, G_2, G_3 \}$,~where $G_3 \mspace{-3mu} = \mspace{-3mu} 13$, and then puncturing $12$ encoded bits equally spaced in the range $\{1,2,\dots,3018\}$. It follows that $d_{\text{IR}} \mspace{-3mu} = \mspace{-3mu} 20$.}
We denote this scheme by $\text{UEPH}_7$.
When our proposed scheme uses sub-optimal encoding matrices as specified in \cite[pg. 484]{nested_conv_2}, then $d_{1} \mspace{-3mu} = \mspace{-3mu} 6$. In such case we denote it by $\text{UEPH}_6$.

We consider two reference schemes which {\em use $\mC_{\text{A}}$ in the first transmission}, and differ in the retransmission.
The first reference scheme, which was presented in \cite{Kallel_1990}, uses the retransmission to {\em equally} protect $\mvec_1$ and $\mvec_2$ via a simple retransmission based on $\mC_{\text{A}}$. This results in a CC $\mC_{\text{K}}(4, 1, 12, 3)$ with the generator polynomials $\{G_1, G_2, G_1, G_2\}$. We denote this scheme by EEPH.
The second reference scheme uses the retransmission to transmit only $\mvec_1$ using $\mC_{\text{IR}}$, thus, applying UEP. As in this scheme $\mvec_2$ are decoded based {\em only} on the first transmission, this scheme is equivalent to separate encoding and is referred to as separate UEPH (SEPUEPH).


\subsection{Simulation Results}
Fig. \ref{fig:BLER} depicts the block error rate (BLER) in decoding $\mvec_1$ vs. $\text{SNR}_{\text{c}}$, for block  Rayleigh fading channel, \textcolor{black}{when only a single retransmission is allowed}, \textcolor{black}{and $10^6$ trials were carried out at each SNR point}.
All the considered schemes apply soft decoding \cite{RCPC}.
Note that all considered schemes {\em use the same CC at the first transmission}, and all require the same number of channel uses. Thus, all the schemes use the same overall transmit energy, and the difference between the performance corresponds to the difference in the code design and in allocating this energy to the IB and SB.
It can be observed that for $\mvec_1$ the BLER of $\text{UEPH}_7$, $\text{UEPH}_6$ and SEPUEPH is the same. This follows as all these schemes retransmit $\mvec_1$ via the same $\mC_{\text{IR}}$. On the other hand, the BLER of EEPH is significantly higher, since in the EEPH scheme, after retransmission, $\mvec_1$ is encoded via $\mC_{\text{K}}$.
Fig. \ref{fig:BER} shows that the BER of $\mvec_2$ for EEPH is significantly lower than the BER achieved by the other schemes. This follows as by combining the first and the second transmissions the EEPH encodes $\mvec_2$ using a CC with $d_{\text{K}} \mspace{-3mu} = \mspace{-3mu} 12$, while the other schemes encode $\mvec_2$ only in the first transmission via a code with significantly smaller free distance.
It can also be observed that both $\text{UEPH}_7$ and  $\text{UEPH}_6$ improve upon SEPUEPH as they use the decoded $\mvec_1$ to assist in decoding $\mvec_2$. Finally, Fig. \ref{fig:BER} demonstrates that using the optimal encoding matrices further improves the~BER.

\section{Conclusion} \label{sec:conc}

We presented a new scheme for UEP-HARQ using pruned CCs.
The benefit of the proposed scheme over existing UEP-HARQ schemes based on CCs is the fact that the codeword corresponding to the SB can be obtained via a projection of the original codeword onto a subcode with a larger free distance. Thus, once the IB are decoded correctly, the projection can be applied and consequently, the BER for the SB is decreased.
\textcolor{black}{This novel concept of designing an HARQ scheme via projection onto a subcode directly extends to general linear codes.}




\end{document}